\documentclass[aps,prc,preprint,tightenlines,groupedaddress,nofootinbib,
showpacs,preprintnumbers,amsmath,amssymb,superscriptaddress]{revtex4}
\usepackage{graphicx}
\usepackage{dcolumn}
\usepackage{bm}
\usepackage{amsmath}
\begin{document}

\title{On three topical aspects of the $N\!=\!28$ isotonic chain}
\author{J. Piekarewicz}
\affiliation{Department of Physics, Florida State 
             University, Tallahassee, FL 32306}
\date{\today} 

\begin{abstract}
The evolution of single-particle orbits along the $N\!=\!28$ isotonic
chain is studied within the framework of a relativistic mean-field
approximation. We focus on three topical aspects of the $N\!=\!28$ 
chain: (a) the emergence of a new magic number at $Z\!=\!14$; (b) the
possible erosion of the $N\!=\!28$ shell; and (c) the weakening of the
spin-orbit splitting among low-$j$ neutron orbits. The present model
supports the emergence of a robust $Z\!=\!14$ subshell gap in
${}^{48}$Ca, that persists as one reaches the neutron-rich isotone
${}^{42}$Si. Yet the proton removal from ${}^{48}$Ca results in a
significant erosion of the $N\!=\!28$ shell in ${}^{42}$Si. Finally,
the removal of $s^{1/2}$ protons from ${}^{48}$Ca causes a
$\sim\!50$\% reduction of the spin-orbit splitting among neutron
$p$-orbitals in ${}^{42}$Si.
\end{abstract}
\pacs{21.10.-k,21.10.Pc,21.60.Jz}
\maketitle 

\section{Introduction}
\label{Sec:Intro}

The study of exotic nuclei is defining a new frontier in nuclear
science. The existent nuclear shell model---the powerful theoretical
edifice build upon insights developed for over half a century from the
study of stable nuclei---may crumble as one leaves the valley of
stability and reaches the extremes of nuclear existence. Fundamental
nuclear-structure concepts, such as the spin-orbit force and the
concomitant emergence of magic numbers, may need to be revised and
redefined~\cite{Nazarewicz:2001,Dobaczewski:2002wp}.

Although the atomic nucleus is shaped by the complex interplay between
strong, electromagnetic, and weak interactions, the nearly 300 stable
nuclei---by their mere existence---probe a relatively small window of
the nuclear chart. Yet the advent of Radioactive Ion Beam Facilities
(RIBFs) have started to change the nuclear landscape and have offered
the first glimpses of exotic new phenomena. One of the central
questions that the new generation of RIBFs will attempt to answer is
what happens to nuclei at the extreme conditions of isospin where the
balance between strong, electromagnetic, and weak interactions differs
significantly from that in stable nuclei. And some answers have
started to emerge. For example, two exotic
nuclei---${}^{42}$Si~\cite{Fridmann:2005} and
${}^{78}$Ni~\cite{Hosmer:2005rm}---have recently been produced at the
National Superconducting Cyclotron of Michigan State University. The
most neutron-rich of the doubly-magic nuclei known to date,
${}^{78}$Ni has a proton-to-neutron ratio of $Z/N\!=\!0.56$,
considerably smaller than the $Z/N\!=\!0.65$ value for the heaviest of
doubly-magic nuclei ${}^{208}$Pb. Further, with twice as many neutrons
as protons---and just one neutron away from the drip
line---${}^{42}$Si reaches a proton-to-neutron fraction of only
$Z/N\!=\!0.5$. For comparison, the heaviest stable member of the
isotopic chain (${}^{30}$Si) has a proton-to-neutron ratio of
$Z/N\!=\!0.88$.

Understanding nuclei at the extreme conditions of isospin will also
shed light on a variety of fundamental questions in astrophysics, such
as the nucleosynthesis of heavy elements and the structure and
dynamics of neutron stars. These phenomena depend critically on
knowledge of the equation of state of neutron-rich matter which at
present is woefully inadequate. For example, in the rapid-neutron
capture process (r-process), stable nuclei (such as iron) are exposed
to a neutron-rich environment that favors neutron-capture timescales
that are considerably shorter than the beta-decay lifetime of the
exotic nuclei that are being produced~\cite{Mathews:1990}. Both the
capture and decay rates depend critically on the shell structure of
nuclei far from stability and in particular on the appearance and
disappearance of magic numbers~\cite{Kratz:1999zd,Thielemann:2001rn}.
Further, neutron-rich nuclei are expected to develop a neutron skin
(or even a neutron halo). Knowledge of the neutron skin of
neutron-rich nuclei places important constraints on the equation of
state of pure neutron matter~\cite{Brown:2000}. Indeed, the slope of
the equation of state of neutron matter---a quantity related to its
pressure---has been shown to be strongly correlated to the neutron
skin of heavy nuclei~\cite{Brown:2000,Furnstahl:2001un} which, in
turn, is correlated to the
structure~\cite{Horowitz:2000xj,Horowitz:2001ya,Carriere:2002bx}, and
cooling~\cite{Horowitz:2002mb} of neutron stars.

Motivated by these topical issues, we have organized the manuscript
around three questions that pertain to the evolution of
single-particle orbits along the $N\!=\!28$ isotonic chain:
(a) do relativistic mean-field models support the emergence of a new
magic number at proton number $Z\!=\!14$? (b) Does the emergence a new
magic number at proton number $Z\!=\!14$ leads to the erosion of the
$N\!=\!28$ neutron shell? (c) Is there a significant weakening of the
spin-orbit splitting among low-$j$ neutron orbits as one moves from
${}^{48}$Ca to ${}^{42}$Si?

The first question is motivated by a recent experiment that suggests
that the neutron-rich nucleus ${}^{42}$Si is doubly magic and has a
(nearly) spherical shape~\cite{Fridmann:2005}. Yet this result is not
without controversy, as the short $\beta$-decay lifetime of ${}^{42}$Si 
has been suggested by some as evidence in favor of a strong 
deformation~\cite{Grevy:2004}.

Theoretically, this issue is equally controversial and strongly
coupled to our second question: is there evidence for the possible
erosion of the $N\!=\!28$ neutron shell at $Z\!=\!14$? On the one
hand, the emergence of a robust subshell closure at $Z\!=\!14$ has
been credited with preventing deformation, thereby lending support to
the notion that ${}^{42}$Si is indeed a new doubly-magic
nucleus~\cite{Retamosa:1996rz,Caurier:2004cq}.  On the other hand, it
has been argued that the progressive removal of $1d^{3/2}$ and $2s^{1/2}$ 
protons causes an erosion of the $N\!=\!28$ neutron shell, resulting 
in a strong deformation in ${}^{42}$Si~\cite{Werner:1994ue,
Werner:1996,Terasaki:1996bf,Lalazissis:1998ew,Peru:2000}. 

The third question explores the possibility of a significant and
systematic weakening of the spin-orbit splitting among the
$2p$-neutron orbitals with proton removal.  The first indication of
the existence of such an effect in the case of ${}^{46}$Ar was
presented in Ref.~\cite{Todd-Rutel:2004tu} where it was shown that in
a relativistic mean-field approach the spin-orbit splitting depends
sensitively on the magnitude of the proton density in the nuclear
interior, and in particular on the occupation of $2s_{1/2}$
orbit. Moreover, it was illustrated that ${}^{46}$Ar departs from the
long-standing paradigm of a spin-orbit potential proportional to the
derivative of the central potential. Since then, this effect has been
confirmed experimentally by Gaudefroy, Sorlin, and collaborators by
using the ${}^{46}Ar(d,p){}^{47}Ar$ transfer reaction in inverse
kinematics~\cite{Gaudefroy:2006}.

A particularly interesting theoretical alternative to the relativistic
mean-field approach presented here has been proposed by Otsuka and
collaborators~\cite{Otsuka:2001nw,Otsuka:2005}. In their approach, the
novel evolution of single particle orbits is attributed to the
spin-isospin structure of the underlying nucleon-nucleon ($NN$)
interaction---particularly the spin-spin~\cite{Otsuka:2001nw} and
tensor~\cite{Otsuka:2005} components. The relativistic mean-field
formulation presented here, based solely on scalar and timelike-vector
Lorentz structures, does not generate any tensor
structure.\footnote{Although the spacelike component of a vector
interaction generates a tensor structure, the spacelike component
vanishes if a spherical ground state is assumed.} Yet, it has been
shown recently by Serra, Otsuka, and collaborators~\cite{Serra:2005},
that a significant contribution from the underlying $NN$ tensor force
manifests itself, through renormalization effects, in the
scalar-isoscalar component of the relativistic mean field potential
(the so-called ``$\sigma$-meson'' contribution). That such
``complicated'' effects emerge in such a simple approach may not come
as a surprise. The strength of an effective field theory framework is
that by fitting a small number of empirical constants to ground-state
data, a variety of correlation effects that go beyond the mean-field
theory get encoded into the parameters of the
model~\cite{Furnstahl:2000in}.  What requires further insights,
however, is the precise role in which the various spin-isospin
structures of the bare $NN$ interaction find their way into the
empirical constants of the effective field theory.

To address the three questions outlined above, the manuscript has 
been organized as follows. In Sec.~\ref{sec:formalism} the effective 
NL3 Lagrangian is introduced~\cite{Lalazissis:1996rd,Lalazissis:1999}
alongside some minor modifications required to describe the physics 
of the $N\!=\!28$ isotonic chain. This effective Lagrangian is used
to compute a variety of ground-state observables at the mean-field 
level. Further details on the implementation of these techniques may
be found in Refs.~\cite{Serot:1984ey,Serot:1997xg,Todd:2003xs}. In
Sec.~\ref{sec:results}, results for the evolution of both neutron and
proton single-particle orbits along the $N\!=\!28$ isotonic chain are
presented and discussed. We conclude in Sec.~\ref{sec:conclusions} with 
a brief summary of our results.

\section{Formalism}
\label{sec:formalism}

The starting point for the calculation of various ground-state properties
is an interacting Lagrangian density of the following form:
\begin{widetext}
\begin{equation}
{\cal L}_{\rm int} =
\bar\psi \left[g_{\rm s}\phi   \!-\!
         \left(g_{\rm v}V_\mu  \!+\!
    \frac{g_{\rho}}{2}{\mbox{\boldmath $\tau$}}\cdot{\bf b}_{\mu}
                               \!+\!
    \frac{e}{2}(1\!+\!\tau_{3})A_{\mu}\right)\gamma^{\mu}
         \right]\psi \nonumber 
                    -
    \frac{\kappa}{3!} (g_{\rm s}\phi)^3 \!-\!
    \frac{\lambda}{4!}(g_{\rm s}\phi)^4 \;.
\label{Lagrangian}
\end{equation}
\end{widetext}
The Lagrangian density includes an isodoublet nucleon field ($\psi$)
interacting via the exchange of two isoscalar mesons, a scalar
($\phi$) and a vector ($V^{\mu}$), one isovector meson ($b^{\mu}$),
and the photon ($A^{\mu}$)~\cite{Serot:1984ey,Serot:1997xg}. In
addition to meson-nucleon interactions the Lagrangian density is
supplemented by two nonlinear meson interactions with coupling
constants denoted by $\kappa$ and $\lambda$. These two terms are 
instrumental in the softening of the equation of state of symmetric 
nuclear matter.

\subsection{Ground-state Properties}
\label{groundstate}

The NL3 parameter set has been enormously successful at describing a
large body of ground-state observables throughout the periodic
table~\cite{Lalazissis:1996rd,Lalazissis:1999}. Yet in the present
contribution a fine tuning of the NL3 parameters is needed to account
for the $1d^{3/2}$-$2s^{1/2}$ proton gap in ${}^{40}$Ca.  Indeed, the
original NL3 parameter set predicts a proton gap of $0.83$~MeV, while
the experimental value has been quoted at
$2.8\!\pm\!0.6$~\cite{Kramer:1990}. To bring the gap closer to the
experimental value an adjustment of the scalar mass $m_{\rm s}$ is
done while maintaining the ratio 
$C_{\rm s}^{2}\!\equiv\!g_{\rm s}^{2}/(m_{\rm s}/M)^{2}/m_{\rm s}^{2}$ 
fixed at $C_{\rm s}^{2}\!\approx\!360$ (see Table~\ref{Table1}). This 
prescription ensures that the binding energy and charge radius of
${}^{40}$Ca remain fixed at (or close to) their experimental values
(see Table~\ref{Table2}). As the main goal of the fine tuning is
simply to obtain a reasonable $1d^{3/2}$-$2s^{1/2}$ proton gap in
${}^{40}$Ca, no attempt was made at re-fitting other parameters of the
model.

\begin{table}
\begin{tabular}{|l||c|c|c|c|c|c|}
 \hline
 Model & $m_{\rm s}$  & $g_{\rm s}^2$ & $g_{\rm v}^2$ & $g_{\rho}^2$
       & $\kappa$ & $\lambda$ \\
 \hline
 \hline
 NL3  & 508.194 & 104.387  & 165.585 & 79.600 & 3.860 & $-$0.016 \\
 NL3  & 450.000 &  82.890  & 165.585 & 79.600 & 3.860 & $-$0.016 \\
\hline
\end{tabular}
\caption{The original NL3 parameter set (first row) and the modified
one (second row) used in the calculations. The modifications are
necessary to (approximately) describe the $1d^{3/2}$-$2s^{1/2}$ proton
gap in ${}^{40}$Ca. The parameter $\kappa$ and the inverse scalar
range $m_{\rm s}$ are given in MeV. The nucleon, omega, and rho masses
are kept fixed at $M\!=\!939$~MeV, $m_{\omega}\!=\!782.5$~MeV, and
$m_{\rho}\!=\!763$~MeV, respectively.}
\label{Table1}
\end{table}

\section{Results}
\label{sec:results}

Having adjusted the NL3 parameter set to reproduce (approximately)
the $1d^{3/2}$-$2s^{1/2}$ proton gap in ${}^{40}$Ca, we are now in 
a position to study the evolution of both proton and neutron
single-particle spectra along the $N\!=\!28$ isotonic chain. 

\subsection{Proton Spectrum}
\label{sec:protons}

\begin{figure}[ht]
\vspace{0.50in}
\includegraphics[height=4in,angle=0]{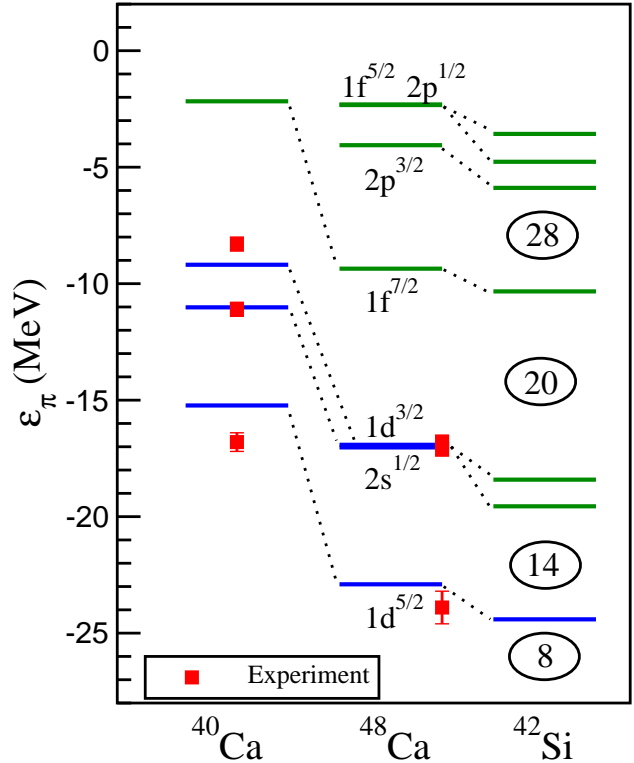}
\caption{(color online) Proton single-particle spectrum in 
         ${}^{40}{\rm Ca}$, ${}^{48}{\rm Ca}$, and
         ${}^{42}{\rm Si}$. Particularly relevant are: (i) the
         near degeneracy of the $1d^{3/2}$-$2s^{1/2}$ in 
         ${}^{48}{\rm Ca}$ and (ii) the emergence of a $Z\!=\!14$ 
         gap in ${}^{48}{\rm Ca}$ that seems to persists in 
         ${}^{42}{\rm Si}$.}
\label{Fig1}
\end{figure}

The evolution of the proton single-particle spectrum as one goes from
${}^{40}_{20}{\rm Ca}_{20}$ to ${}^{42}_{14}{\rm Si}_{28}$ via
${}^{48}_{20}{\rm Ca_{28}}$ is depicted in Fig.~\ref{Fig1}.  The same
information (limited to the $s\!-\!d$ shell) alongside binding
energies per nucleon and root-mean-square charge radii is listed in
Table~\ref{Table2}. The figure is particularly suggestive of a new
shell closure at proton number $Z\!=\!14$ that emerges from the
near-degeneracy between the $2s^{1/2}$ and the $1d^{3/2}$ proton
orbitals at neutron number $N\!=\!28$~\cite{Cottle:1998}.  This
critical feature is well accounted for by the relativistic mean-field
model. Indeed, the energy gap between the $1d^{3/2}$ and $2s^{1/2}$
proton orbitals is reduced from $\sim\!2$~MeV in ${}^{40}{\rm Ca}$ to
a mere $110$~keV in ${}^{48}$Ca.  This results in a robust $6$~MeV
energy gap between the $1d^{5/2}$ orbital and the quasi-degenerate
pair. This quasi-degeneracy gets only slightly diluted by stripping
${}^{48}$Ca from its six valence protons. In this form, one reaches
the ``doubly-magic'' nucleus ${}^{42}$Si, with a predicted proton gap
at the Fermi surface of almost $5$~MeV. (Note that the original NL3
parameter set, although accurately calibrated, predicts a
$1d^{3/2}$-$2s^{1/2}$ proton gap in ${}^{48}$Ca of $-1.8$~MeV.)

\begin{table}
\begin{tabular}{|c||c|c|c|c|c|}
 \hline
  Nucleus & $B/A$~(MeV) & $r_{\rm ch}$~(fm) 
          & $\epsilon(1d^{5/2})$~(MeV) 
          & $\epsilon(2s^{1/2})$~(MeV) 
          & $\epsilon(1d^{3/2})$~(MeV) \\ 
 \hline
  ${}^{40}$Ca & 8.55 & 3.52 & $-15.23~(6.04)$ 
              & $-11.02~(1.83)$ & $ -9.19~(0.00)$ \\
  ${}^{48}$Ca & 8.71 & 3.51 & $-22.91~(5.99)$ 
              & $-17.03~(0.11)$ & $-16.92~(0.00)$ \\
  ${}^{42}$Si & 7.41 & 3.27 & $-24.41~(6.00)$ 
              & $-19.55~(1.14)$ & $-18.41~(0.00)$ \\
\hline
\end{tabular}
\caption{Evolution of the the proton single-particle spectrum in
         the $s\!-\!d$ shell. Also shown are binding energies per 
         nucleon and charge radii. Quantities in parenthesis are 
         single-proton energies relative to the $\epsilon(1d^{3/2})$.} 
\label{Table2}
\end{table}

\begin{figure}[ht]
\vspace{0.50in}
\includegraphics[width=4.5in,angle=0]{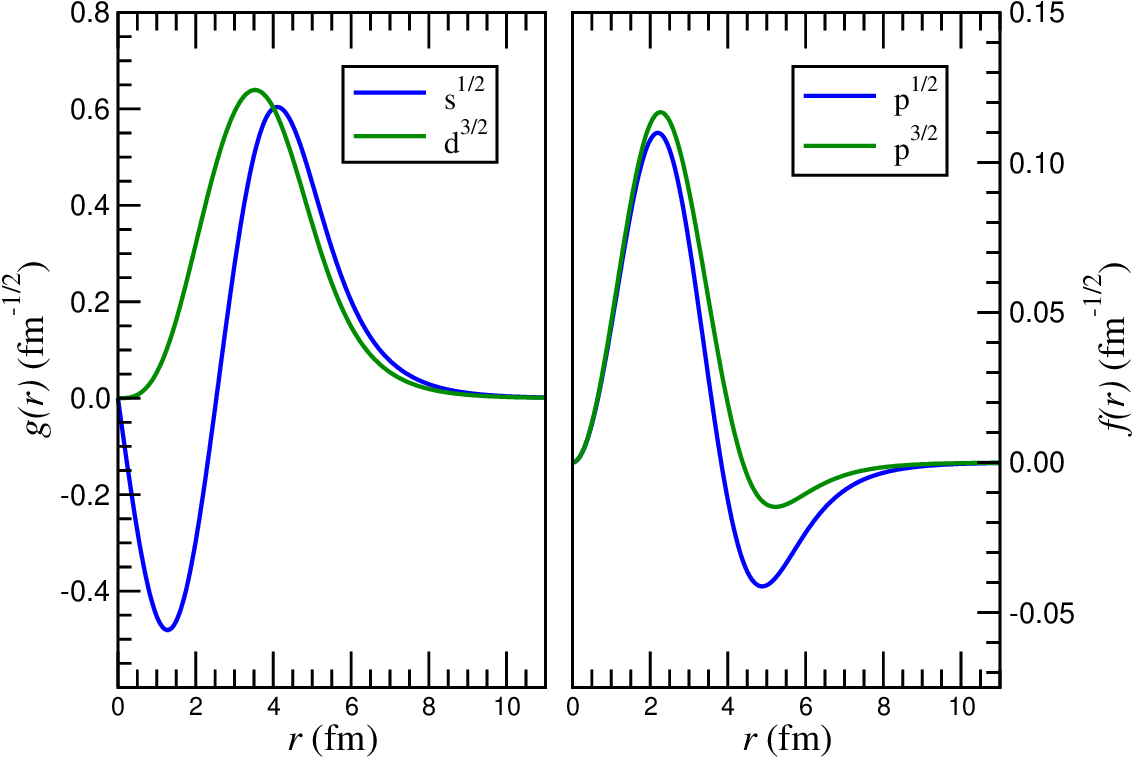}
\caption{(color online) Upper (left-hand panel) and lower (right-hand 
         panel) components of the $2s^{1/2}$ and $1d^{3/2}$ proton 
	 orbitals in ${}^{48}$Ca. The left-hand panel illustrates the 
	 different character of the upper components in the nuclear 
	 interior. In contrast, the right-hand panel accentuates the 
	 similarity among the corresponding lower components and provides 
	 a natural explanation for the appearance of pseudospin symmetry 
	 in nuclei.}
\label{Fig2}
\end{figure}

The quasi-degeneracy of the $2s^{1/2}$ and the $1d^{3/2}$ orbitals,
the main physics behind the appearance of a new magic number at
$Z\!=\!14$, is related to the old phenomenon of ``pseudospin''
symmetry.  This is a symmetry that manifests itself among
$(n+1)lj\!=\!l\!+\!1/2$ and $n(l+2)j\!=\!l\!+\!3/2$ orbitals, with
$(n\!-\!1)$ the number of nodes, $l$ the orbital angular momentum, and
$j$ the total angular momentum of the single-particle orbital. Nothing
a-priori suggests a symmetry among pseudospin-orbit partners ({\it
e.g.,} $2s^{1/2}$ and $1d^{3/2}$ orbitals), as their wave functions
differ in both angular momentum and number of nodes, a feature that is
clearly illustrated on the left-hand panel of Fig.~\ref{Fig2}. Yet
pseudo-spin symmetry appears to repair these differences at large
separations ($r\!\gtrsim\!5$~fm), leading to an exponential behavior
controlled by energies that differ by only 110~keV. We note that while
the quantitative value of the $1d^{3/2}$-$2s^{1/2}$ energy splitting
is model dependent, their near-degeneracy is a robust result that is
well described within the relativistic approach.

In a series of seminal papers, Ginnochio and collaborators have shown
that pseudospin symmetry appears naturally in a relativistic approach
as a consequence of the similarity among the ``small'' (or lower)
components of the Dirac orbitals, rather than from a similarity among
the dominant upper components. In the particular case of the
$2s^{1/2}$ and $1d^{3/2}$ pseudo-spin orbit partners, the lower
components carry quantum numbers given by $p^{1/2}$ and $p^{3/2}$,
respectively. Thus, both orbitals experience the same centrifugal
barrier. Further, because of their respective boundary conditions,
they also display the same number of
nodes~\cite{Piekarewicz:1993se}. The only difference among these
orbitals arises from the presence of a ``small'' 
{\it pseudospin-orbit} potential that results from the cancellation of
large scalar and vector potentials. Note that large scalar and vector
potentials are the hallmark of all successful relativistic mean-field
theories and the main reason behind the large {\it spin-orbit}
potential, where the large potentials add rather than cancel.  
The uncanny resemblance among the lower components of the $2s^{1/2}$ 
and $1d^{3/2}$ Dirac spinors is depicted on the right-hand panel of
Fig.~\ref{Fig2}. The role of pseudospin symmetry in exotic nuclei is 
an interesting topic that deserves further consideration.


\subsection{Neutron Spectrum}
\label{sec:neutrons}

In the previous section the evolution of the proton spectrum with 
increasing neutron number was explored. It was suggested that the 
filling of the $1f^{7/2}$ neutron orbital in the Calcium isotopes 
resulted in the appearance of a new magic number at $Z\!=\!14$ that 
appears to persists as one removes the six valence protons from 
${}^{48}$Ca, thereby reaching the ``doubly-magic'' nucleus 
${}^{42}_{14}{\rm Si}_{28}$~\cite{Retamosa:1996rz,Caurier:2004cq,
Fridmann:2005}. 

\begin{figure}[ht]
\vspace{0.50in}
\includegraphics[height=4in,angle=0]{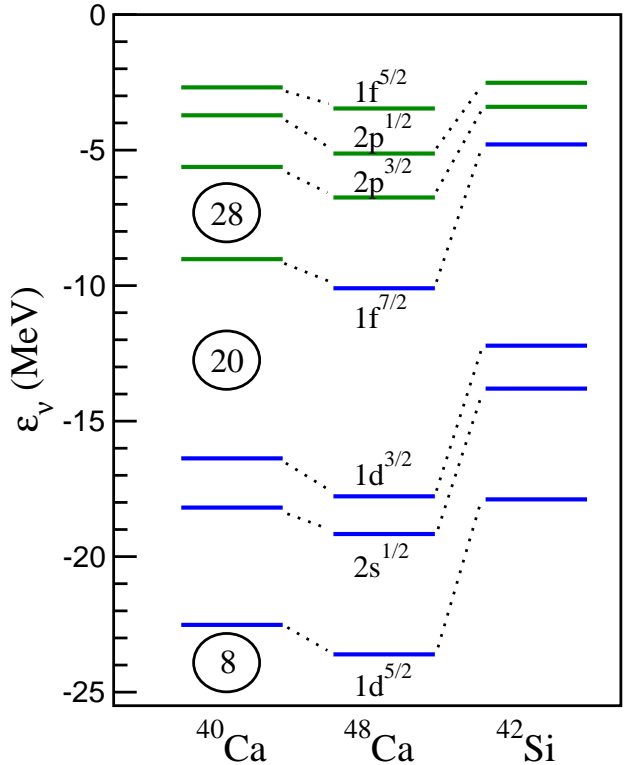}
\caption{(color online) Neutron single-particle spectrum in 
         ${}^{40}{\rm Ca}$, ${}^{48}{\rm Ca}$, and
         ${}^{42}{\rm Si}$. Particularly relevant are: (i) the
	 return of the $1f^{7/2}$ orbital to its parent shell
         in ${}^{42}{\rm Si}$ and (ii) the $\sim\!50$\% reduction 
         in the spin-orbit splitting of the $p$-orbitals in 
         ${}^{42}{\rm Si}$ relative to ${}^{48}{\rm Ca}$.}
\label{Fig3}
\end{figure}

In this section we are interested in exploring a complementary effect,
namely, the evolution of the neutron single-particle spectrum in the 
$N\!=\!28$ isotones as a function of the progressive removal of the 
six valence protons in ${}^{48}$Ca. Thus, we start this section by
displaying in Fig.~\ref{Fig3} the neutron counterpart to the proton
single-particle spectrum depicted in Fig.~\ref{Fig1}. Particularly
interesting is the prediction that as new magic number emerges at
proton number $Z\!=\!14$, an old one disappears at neutron number
$N\!=\!28$. While this result is at variance with the conclusions of
Refs.~\cite{Retamosa:1996rz,Caurier:2004cq}, it supports the findings
of Refs.~\cite{Werner:1994ue,Werner:1996,Terasaki:1996bf,
Lalazissis:1998ew,Peru:2000}.  In the present relativistic mean-field
model the gradual return of the $1f^{7/2}$-orbit to its parent
$2p\!-\!1f$ shell is caused by the repulsive vector-isovector
interaction.  As the six protons are removed from the
$2s^{1/2}$-$1d^{3/2}$ orbitals, the isovector interaction---which is
driven by the difference between proton and neutron vector
densities---becomes increasingly repulsive for neutrons, resulting in
a $\sim\!2$~MeV erosion of the $N\!=\!28$ gap (see
Table~\ref{Table3}). Such a significant reduction is likely to cause a
redistribution of single-particle strength among the various neutron
orbitals in the $2p\!-\!1f$ shell. This could become an important
source of soft positive-parity ({\it e.g.,} quadrupole) excitations
and/or deformation. Indeed, a preliminary random-phase-approximation
(RPA) calculation of quadrupole strength based on the formalism
outlined in Refs.~\cite{Piekarewicz:2000nm,Piekarewicz:2001nm}
reveals a low-lying $2^{+}$ excitation at $\sim\!0.5$~MeV in
${}^{42}$Si.

The one aspect that remains to be addressed is the quenching of the
spin-orbit splitting of the neutron $p$-orbitals as a function of
proton $s^{1/2}$ removal.  In Ref.~\cite{Todd-Rutel:2004tu} it was
proposed that the depletion of $s^{1/2}$ strength could have a
dramatic effect on the spin-orbit splitting of low angular momentum
orbitals. In particular, it was demonstrated that the spin-orbit
splitting of the $p^{3/2}$-$p^{1/2}$ neutron orbitals depends
sensitively on the occupation of $s^{1/2}$ proton orbits. Two exotic
nuclei were identified --- ${}^{46}$Ar and ${}^{206}$Hg --- in which
the depletion of $s^{1/2}$ proton strength yields a dramatic quenching
in the spin-orbit splitting of neutron $p$-orbitals near the Fermi
surface. Since then, this prediction has been confirmed
experimentally.  Indeed, a significant weakening of the spin-orbit
splitting among the $2p^{3/2}$-$2p^{1/2}$ neutron orbitals in
${}^{46}$Ar has been reported recently by Gaudefroy, Sorlin, and
collaborators~\cite{Gaudefroy:2006}. We now demonstrate that the
weakening of the spin-orbit splitting continues as one reaches 
${}^{42}$Si.

\begin{figure}[ht]
\vspace{0.50in}
\includegraphics[width=4.5in,angle=0]{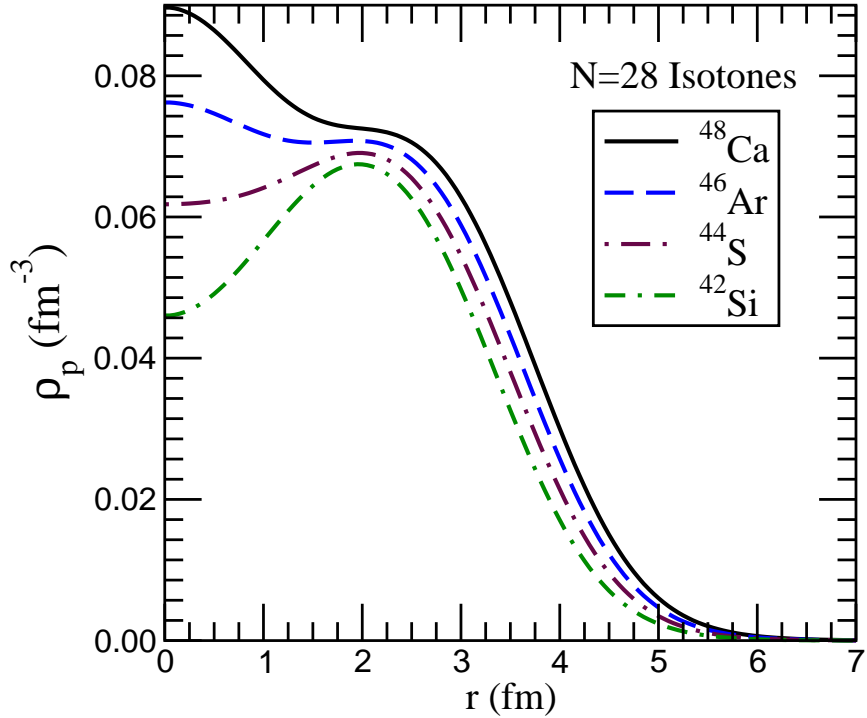}
\caption{(color online) Point proton densities for the four $N\!=\!28$
         isotones considered in the text. The development of a proton
         ``hole'' in the nuclear interior is responsible for the
         weakening of the spin-orbit splitting among the neutron
         $p$-orbitals.}
\label{Fig4}
\end{figure}

The weakening of the spin-orbit splitting among the neutron
$p$-orbitals is clearly discernible in Fig.~\ref{Fig3}. The removal of
the $2s^{1/2}$ protons is responsible for ``carving'' a hole in the
nuclear interior that dramatically reshapes the spin-orbit potential,
leaving a strong imprint on neutron orbitals of low angular
momentum~\cite{Todd-Rutel:2004tu}.  The development of a proton hole
in the nuclear interior as $2s^{1/2}$ protons are progressively
removed is nicely illustrated in Fig.~\ref{Fig4}. Note that a uniform
filling fraction of $2/3$ and $1/3$ has been assumed for ${}^{46}$Ar
and ${}^{44}$S, respectively. This choice is motivated by the
quasi-degeneracy of the $1d^{3/2}$-$2s^{1/2}$ orbitals.

\begin{figure}[ht]
\vspace{0.50in}
\includegraphics[width=4.5in,angle=0]{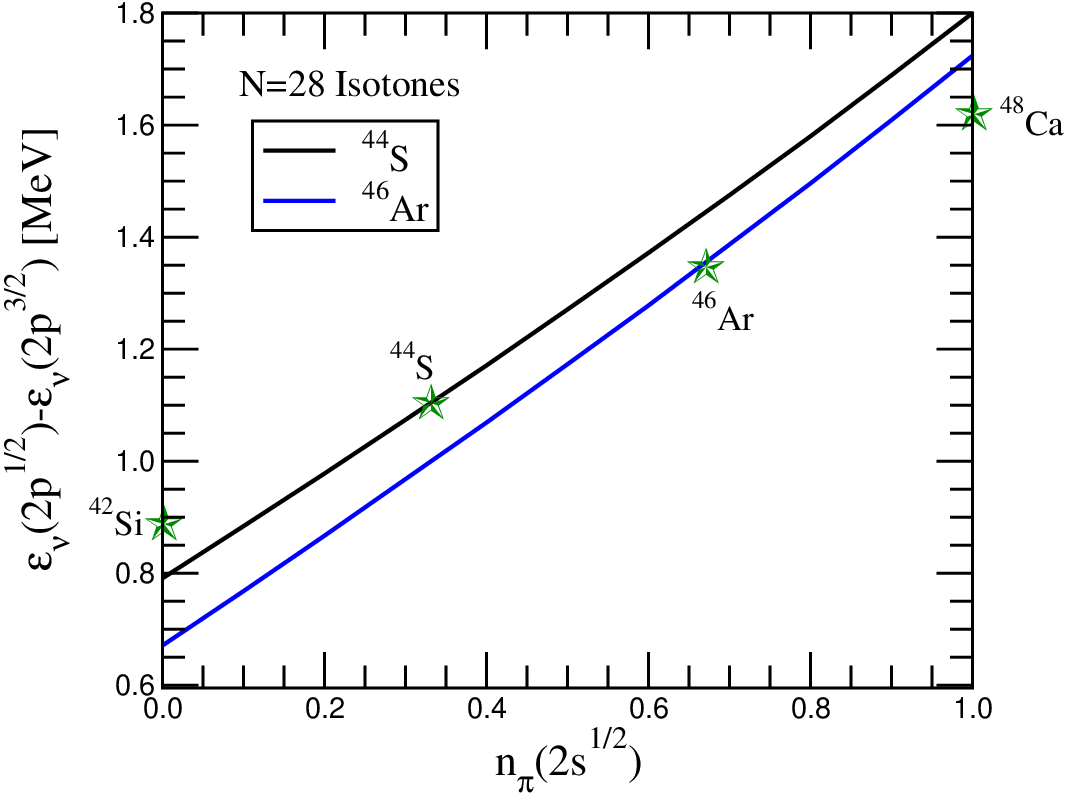}
\caption{Spin-orbit splitting of the neutron $p$-orbitals as a 
         function of the proton $2s^{1/2}$ occupancy for the four 
         $N\!=\!28$ isotones considered in the text. Values indicated 
         with a ``star'' for ${}^{44}$S and ${}^{46}$Ar have been 
	 obtained by assuming a uniform occupancy of the 
	 quasi-degenerate $2s^{1/2}$-$1d^{3/2}$ proton orbitals.}
\label{Fig5}
\end{figure}

In Fig.~\ref{Fig5} we display the sensitivity of the spin-orbit
splitting among neutron $p$-orbitals to the filling fraction of the
proton $2s^{1/2}$ orbit. The removal of $2s^{1/2}$ strength yields a
sharp increase in the magnitude of the spin-orbit interaction in the
nuclear interior that cancels part of the surface contribution. One
can observe that in the case that only $1d^{3/2}$ protons are removed
[$n_{\pi}(2s^{1/2})\!\equiv\!1$] the model actually predicts a slight
increase in the spin-orbit splitting. Yet the more plausible
assumption of a uniform proton removal from the quasi-degenerate
orbits leads to a spin-orbit quenching of about 20\% and 30\% for
${}^{46}$Ar and ${}^{44}$S, respectively. These values are indicated
as stars in the figure. In the case of ${}^{42}$Si, where the
$2s^{1/2}$ removal is complete, the spin-orbit quenching reaches
almost 50\%.  Note that while the quenching of the spin-orbit
splitting is a robust prediction of the relativistic mean-field
model---as the interior contribution to the spin-orbit potential
always works against the dominant surface contribution---the precise
magnitude of the quenching is model dependent. For example, the
original NL3 parameter set predicts a $2p^{1/2}$-$2p^{3/2}$ splitting
in ${}^{46}$Ar at zero $2s^{1/2}$ filling of only
$0.17$~MeV~\cite{Todd-Rutel:2004tu}, rather than the $0.67$~MeV value
obtained here with the slightly modified parameter set.

\begin{table}
\begin{tabular}{|c||c|c|c|}
 \hline
  Nucleus & $\epsilon(1f^{7/2})$~(MeV) 
          & $\epsilon(2p^{3/2})$~(MeV) 
          & $\epsilon(2p^{1/2})$~(MeV) \\ 
 \hline
  ${}^{48}$Ca & $-10.10~(4.97)$ & $-6.75~(1.62)$ & $-5.13~(0.00)$ \\
  ${}^{46}$Ar & $ -8.35~(4.13)$ & $-5.57~(1.35)$ & $-4.22~(0.00)$ \\
  ${}^{44}$S  & $ -6.58~(3.24)$ & $-4.45~(1.11)$ & $-3.34~(0.00)$ \\
  ${}^{42}$Si & $ -4.79~(2.27)$ & $-3.41~(0.89)$ & $-2.52~(0.00)$ \\
\hline
\end{tabular}
\caption{Evolution of the neutron $p\!-\!f$ shell with proton
         removal from the $1d^{3/2}$ and $2s^{1/2}$ orbitals.
         Quantities in parenthesis are single-neutron energies 
         relative to the $\epsilon(2p^{1/2})$.} 
\label{Table3}
\end{table}

\section{Conclusions}
\label{sec:conclusions}

A study of the $N\!=\!28$ isotonic chain was conducted within the
framework of the relativistic mean-field approximation. For our
calculations, we relied on a slightly modified NL3 parameter
set~\cite{Lalazissis:1996rd}. While accurately calibrated and highly
successful in describing a variety of ground state
observables~\cite{Lalazissis:1999}, the original NL3 set
underestimates the $1d^{3/2}$-$2s^{1/2}$ proton gap ($\sim\!800$~keV
{\it vs} $\sim\!2$~MeV) in ${}^{40}{\rm Ca}$.  Thus, a fine tuning of
the NL3 parameter set was performed to properly account for the
gap. The adjustments were done in such a way that the binding energy
and charge radius of ${}^{40}$Ca remain close to their experimental
values. This adjustment, alone and with nothing else, yields a nearly
degenerate $1d^{3/2}$-$2s^{1/2}$ proton pair and a robust $Z\!=\!14$
gap in ${}^{48}$Ca, in agreement with experiment.

The aim of the present study was threefold. First, to determine if the
appearance of a new magic number at $Z\!=\!14$ in ${}^{48}$Ca persists
as one reaches ${}^{42}$Si~\cite{Fridmann:2005}. Second, to explore the 
persistence---or lack-thereof---of the $N\!=\!28$ shell closure in 
${}^{42}$Si~\cite{Werner:1994ue,Werner:1996,Retamosa:1996rz,
Lalazissis:1998ew,Cottle:1998,Peru:2000}. Finally, to monitor the
weakening of the spin-orbit splitting of the $2p^{3/2}$-$2p^{1/2}$
neutron orbits as a function of the occupancy of the $2s^{1/2}$ 
proton orbital~\cite{Todd-Rutel:2004tu,Gaudefroy:2006}.

Regarding the first topic, we found that the addition of 8 neutrons to
the $1f^{7/2}$ orbit reduces the $1d^{3/2}$-$2s^{1/2}$ proton gap from
$\sim\!2$~MeV in ${}^{40}$Ca to a mere 110~keV in ${}^{48}{\rm
Ca}$. This, in turn, yields a robust $\sim\!6$~MeV energy gap at
proton number $Z\!=\!14$. As the removal of all 6 protons from the
degenerate $1d^{3/2}$-$2s^{1/2}$ orbitals weakens the $Z\!=\!14$ gap
only slightly (by $\sim\!1$~MeV), our results support the emergence of
a well developed proton subshell closure at
$Z\!=\!14$~\cite{Fridmann:2005}.

Yet the present relativistic mean-field model predicts that as protons
are progressively removed from the $1d^{3/2}$-$2s^{1/2}$ orbitals, the
$1f^{7/2}$ neutron orbit returns to its parent $fp$-shell---leading to
the disappearance of the magic number $N\!=\!28$.  In the present
model the $N\!=\!28$ gap is systematically reduced from $3.4$~MeV in
${}^{48}$Ca, to $2.8$~MeV in ${}^{46}$Ar, to $2.1$~MeV in ${}^{44}$S,
and ultimately to $1.4$~MeV in ${}^{42}$Si. The significant reduction
in the gap is caused by the Lorentz structure of the isovector
interaction, assumed here to be of vector character. For neutron-rich
nuclei, a vector-isovector structure generates repulsion for the
majority species (neutrons) and attraction for the minority species
(protons). Thus, in the present model the proton removal is ultimately
responsible for the return of the $1f^{7/2}$ neutron orbit to its
parent shell. It is conceivable that the disappearance of the
$N\!=\!28$ magic number could promote both deformation and/or soft
quadrupole excitations. Indeed, some of our preliminary RPA
calculations suggest a low-energy quadrupole excitation at around
$500$~keV. Yet the ultimate shape of ${}^{42}$Si---either
spherical~\cite{Retamosa:1996rz,Cottle:1998,Fridmann:2005} or
deformed~\cite{Werner:1996,Lalazissis:1998ew,Peru:2000}---its
single-particle structure, and the character of its low-energy
excitations, will continue to be the source of considerable debate for
years to come.

Finally, the weakening of the spin-orbit splitting among low-$j$
neutron orbits---an effect first predicted for $p$-orbitals in
${}^{46}$Ar and ${}^{206}$Hg~\cite{Todd-Rutel:2004tu}, and recently
confirmed experimentally in ${}^{46}$Ar~\cite{Gaudefroy:2006}---is
strongly correlated to the occupancy of the proton $2s^{1/2}$
orbital. Indeed, the removal of $s^{1/2}$ strength induces significant
structure in the nuclear interior that is responsible for canceling
part of the dominant surface contribution to the spin-orbit
potential. This cancellation yields a significant quenching of 
$\sim\!50$\% in the spin-orbit splitting of the $2p$ neutron orbitals 
in ${}^{42}$Si relative to the corresponding splitting in ${}^{48}$Ca.

\begin{acknowledgments}
The author gratefully acknowledges Professors P. Cottle, O. Sorlin,
A. Volya, and I. Wiedenhover for valuable discussions and illuminating
insights. This work was supported in part by DOE grant
DE-FG05-92ER40750.
\end{acknowledgments}

\vfill\eject
\bibliography{ReferencesJP}

\end{document}